\documentclass[aps]{revtex4}
\usepackage{amsfonts}
\usepackage{amsmath}
\usepackage{amssymb,epsf}

\begin{document}

\title{Magnetic branes in Gauss-Bonnet gravity with nonlinear
electrodynamics: \\
correction of magnetic branes in Einstein-Maxwell gravity}
\author{Seyed Hossein Hendi$^{1,2}$\footnote{email address: hendi@shirazu.ac.ir}, Shahram Panahiyan$^{1}$\footnote{
email address: ziexify@gmail.com} and Behzad Eslam
Panah$^{1}$\footnote{email address:
behzad$_{-}$eslampanah@yahoo.com}} \affiliation{$^1$ Physics
Department and Biruni Observatory, College of Sciences, Shiraz
University, Shiraz 71454, Iran\\
$^2$ Research Institute for Astronomy and Astrophysics of Maragha (RIAAM),
Maragha, Iran}

\begin{abstract}
In this paper, we are considering two first order corrections to both
gravity and gauge sides of the Einstein-Maxwell gravity: Gauss-Bonnet
gravity and quadratic Maxwell invariant as corrections. We obtain
horizonless magnetic solutions by implying a metric which representing a
topological defect. We analyze the geometric properties of the solutions and
investigate the effects of both corrections, and find that these solutions
may be interpreted as the magnetic branes. We study the singularity
condition and find a nonsingular spacetime with a conical geometry. We also
investigate the effects of different parameters on deficit angle of
spacetime near the origin.
\end{abstract}

\maketitle

\section{Introduction}

Magnetic strings/branes may be interpreted as topological defects
that were formed during phase transition of the early universe
\cite{Phase}. These defects contain information regarding early
structure of the universe and also its evolution \cite{velkin}. On
the other hand, considering the AdS/CFT correspondence, these
horizonless magnetic solutions in presence of negative
cosmological constant may contain information regarding a quantum
field theory on the boundary of the AdS spacetime \cite{Ads}.
These topological defects have wide variety of applications in
quantum gravity and has been studied in different context such as
hadron dynamics \cite{hadron}, anti-ferromagnetic crystals
\cite{crystal}, Yang-Mills plasma \cite{plasma} and also, in
studying quantum criticality \cite{kraus}. These magnetic
branes/strings have been studied through some papers and their
properties for different cases of gravity and nonlinear
electromagnetic fields have been derived \cite{Mag}. Motivated by
these facts, we study magnetic branes in presence of two mentioned
corrections and investigate the effects of these two corrections
on properties of the magnetic branes.

From the other point of view, Einstein (EN) gravity has been a
successful theory for describing many phenomena, whereas in some
aspects it confronts some fundamental problems \cite{Stelle}. In
order to overcome these problems, alternative theories of gravity
or generalization of the EN gravity have been introduced
\cite{Lovelock,Brans,FR}. One of these theories is generalization
of the EN gravity to the well-known Gauss-Bonnet (GB) theory. This
generalization solves some of shortcomings of the EN gravity and
gives a renewed view and properties in gravitational context
\cite{GB}. This theory of gravity has been studied for different
astrophysical objects \cite{GB1}. The properties of the GB
gravity, may attract one to the idea that not to consider the GB
gravity as a generalization, but as a correction to EN gravity. In
other words, one can consider first order of the GB parameter,
$\alpha $, as a correction and study its effects on the properties
of solutions. This fact shows that one can take small values of
the GB parameter into account and interpret the effects of the GB
correction as a perturbation to EN gravity. This consideration
enables us to study the effects of the GB parameter in more
details. In this paper, we consider the GB gravity not as a
generalization but as a correction or perturbation to EN theory
\cite{HendiP}.

Naturally, most of the systems that we are studying are nonlinear
or they have nonlinear properties. In order to have more realistic
results, one should take into account the nonlinear behavior of
these systems. Maxwell theory of electrodynamics is a linear
theory which works well in many aspects but fails regarding some
important issues. In order to overcome its problems, different
theories of nonlinear electrodynamics (NED) were introduced
\cite{Nonlinear}. Among them, Born-Infeld (BI) type ones are quite
interesting due to their properties and the fact that these
theories
may arise from low energy limits of the effective string theory \cite{String}%
. For large values of nonlinearity parameter, these BI type
theories lead to same behavior such as Maxwell theory. In order to
avoid complexity that nonlinear electromagnetic fields pose, one
can consider the effects of nonlinearity as a correction to the
Maxwell field. In other words, one can add the quadratic Maxwell
invariant to the Lagrangian of the Maxwell theory
\cite{HendiP,Hendi2,Hendi3} to obtain NED as a correction. On the
other hand, it is arguable that in order to obtain physical
results that are consistent with experiments, one should consider
the weak effects of nonlinearity. Using the series expansion of BI
type theories in weak field limit, one finds that the first term
is Lagrangian of the Maxwell theory and the second term is
proportional to the quadratic Maxwell invariant \cite{Hendi1}.
Therefore, in this paper, we are considering quadratic Maxwell
invariant as a correction of the Maxwell electrodynamics and study
its effects on the properties of solution.

The structure of the paper will be as follow. In next section, we will
present fields equation and obtain metric function for the case of the
magnetic branes. We will plot some graphs in order to study the effects of
corrections on metric function. Also we will study the geometrical structure
of obtained solutions and investigate the effects of different parameters on
deficit angle and conical structure of the magnetic branes through graphs.
Last section is devoted to closing remarks.

\section{Static solutions}

In order to study horizonless magnetic branes, we consider the
following metric for $d$-dimensions \cite{Dias}
\begin{equation}
ds^{2}=-\frac{\rho ^{2}}{l^{2}}dt^{2}+\frac{d\rho ^{2}}{f(\rho )}%
+l^{2}f(\rho )d\phi ^{2}+\frac{\rho ^{2}}{l^{2}}dX^{2},  \label{Metric1}
\end{equation}%
where $l$ is a scale factor related to the cosmological constant,
$f(\rho)$ is a function of coordinate $\rho$ and
$dX^{2}=\sum_{i=1}^{d-3}dx_{i}^{2}$ is the Euclidean metric on the
$(d-3)$-dimensional submanifold. The angular coordinate $\phi $ is
dimensionless and ranges in $[0,2\pi ]$, while $x_{i}$ ranges in
$(-\infty ,\infty )$. Due to fact that we are interested in
solutions that contain the GB gravity and a correction to Maxwell
field, we consider the following field equations
\cite{HendiP,Hendi2}:
\begin{equation}
\partial _{\mu }\left( \sqrt{-g}\mathcal{L}_{\mathcal{F}}F^{\mu \nu }\right)
=0,  \label{FE1}
\end{equation}%
\begin{equation}
\Lambda g_{\mu \nu }+G_{\mu \nu }^{(1)}+\alpha G_{\mu \nu }^{(2)}+O\left(
\alpha ^{2}\right) =\frac{1}{2}g_{\mu \nu }\mathcal{L}(\mathcal{F})-2%
\mathcal{L}_{\mathcal{F}}F_{\mu \lambda }F_{\nu }^{\lambda },  \label{FE2}
\end{equation}%
where $\mathcal{L}_{\mathcal{F}}=\frac{d\mathcal{L}(\mathcal{F})}{d\mathcal{F%
}}$, in which $\mathcal{L}(\mathcal{F})$ is the Lagrangian of NED; $\Lambda
=-\frac{(d-1)(d-2)}{2l^{2}}$ and $G_{\mu \nu }^{(1)}=R_{\mu \nu }-\frac{1}{2}%
g_{\mu \nu }R$ are, respectively the cosmological constant and the EN
tensor; $\alpha $ is the GB coefficient and
\begin{equation}
G_{\mu \nu }^{(2)}=2(R_{\mu \sigma \kappa \tau }R_{\nu }^{\phantom{\nu}%
\sigma \kappa \tau }-2R_{\mu \rho \nu \sigma }R^{\rho \sigma }-2R_{\mu
\sigma }R_{\phantom{\sigma}\nu }^{\sigma }+RR_{\mu \nu })-\frac{\mathcal{L}%
^{(2)}}{2}g_{\mu \nu },
\end{equation}%
where $\mathcal{L}^{(2)}$ denotes the Lagrangian of the GB gravity, given as
\begin{equation}
\mathcal{L}^{(2)}=R_{\mu \nu \gamma \delta }R^{\mu \nu \gamma \delta
}-4R_{\mu \nu }R^{\mu \nu }+R^{2}.  \label{L2}
\end{equation}

We consider the following Lagrangian for the electromagnetic field \cite%
{HendiP,Hendi2,Hendi3}
\begin{equation}
\mathcal{L}(\mathcal{F})=-\mathcal{F+}\beta \mathcal{F}^{2}+O\left( \beta
^{2}\right) ,  \label{L(F)}
\end{equation}%
where $\beta $ is the nonlinearity parameter and the Maxwell invariant $%
\mathcal{F}=F_{\mu \nu }F^{\mu \nu }$, in which $F_{\mu \nu }=\partial _{\mu
}A_{\nu }-\partial _{\nu }A_{\mu }$ is the electromagnetic field tensor and $%
A_{\mu }$ is the gauge potential. It is easy to show that the electric field
comes from the time component of the vector potential ($A_{t}$), while the
magnetic field is associated with the angular component ($A_{\phi }$). The
black hole solutions of GB gravity in presence of this nonlinear
electromagnetic field were obtained previously \cite{HendiP}. In this paper
we are looking for horizonless solutions with conical singularity which are
not interpreted as black holes but magnetic branes solutions. Since we are
looking for the magnetic solutions, we consider the following form of gauge
potential
\begin{equation}
A_{\mu }=h(\rho )\delta _{\mu }^{\phi }.  \label{Amu}
\end{equation}

Using Eq. (\ref{Amu}) with the mentioned NED, one can show that the
electromagnetic field equation, (\ref{FE1}), reduces to the following
differential equation
\begin{equation}
\left[ \left( d-2\right) E+E^{\prime }\rho \right] l^{2}+4E^{2}\left[ \left(
d-2\right) E+3E^{\prime 2}\rho \right] \beta +O(\beta ^{2})=0,  \label{FE11}
\end{equation}%
where prime denotes the first derivative with respect to $\rho $ and $%
E=h^{\prime }(\rho )$. Solving Eq. (\ref{FE11}), one obtains
\begin{equation}
E(\rho )=\frac{2ql^{2}}{\rho ^{d-2}}-\frac{32q^{3}l^{4}\beta }{\rho
^{3\left( d-2\right) }}+O\left( \beta ^{2}\right) ,  \label{h(rho)}
\end{equation}%
where $q$ is an integration constant related to the electric charge. We
should note that for small values of $\beta $ all relations reduce to the
corresponding relations of the Maxwell theory.

\begin{figure}[tbp]
$%
\begin{array}{ccc}
\epsfxsize=5.6cm\epsffile{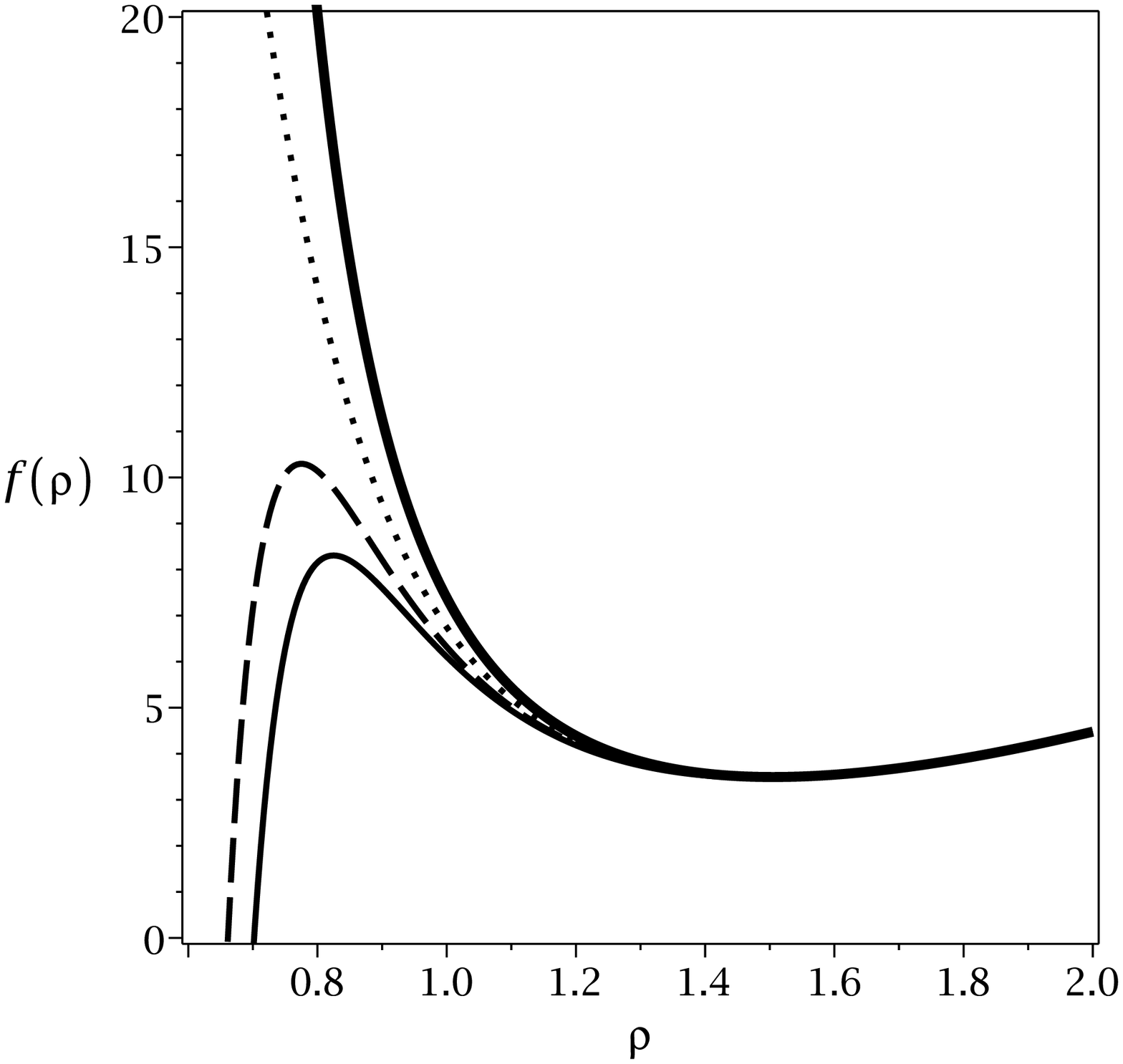} & %
\epsfxsize=5.6cm\epsffile{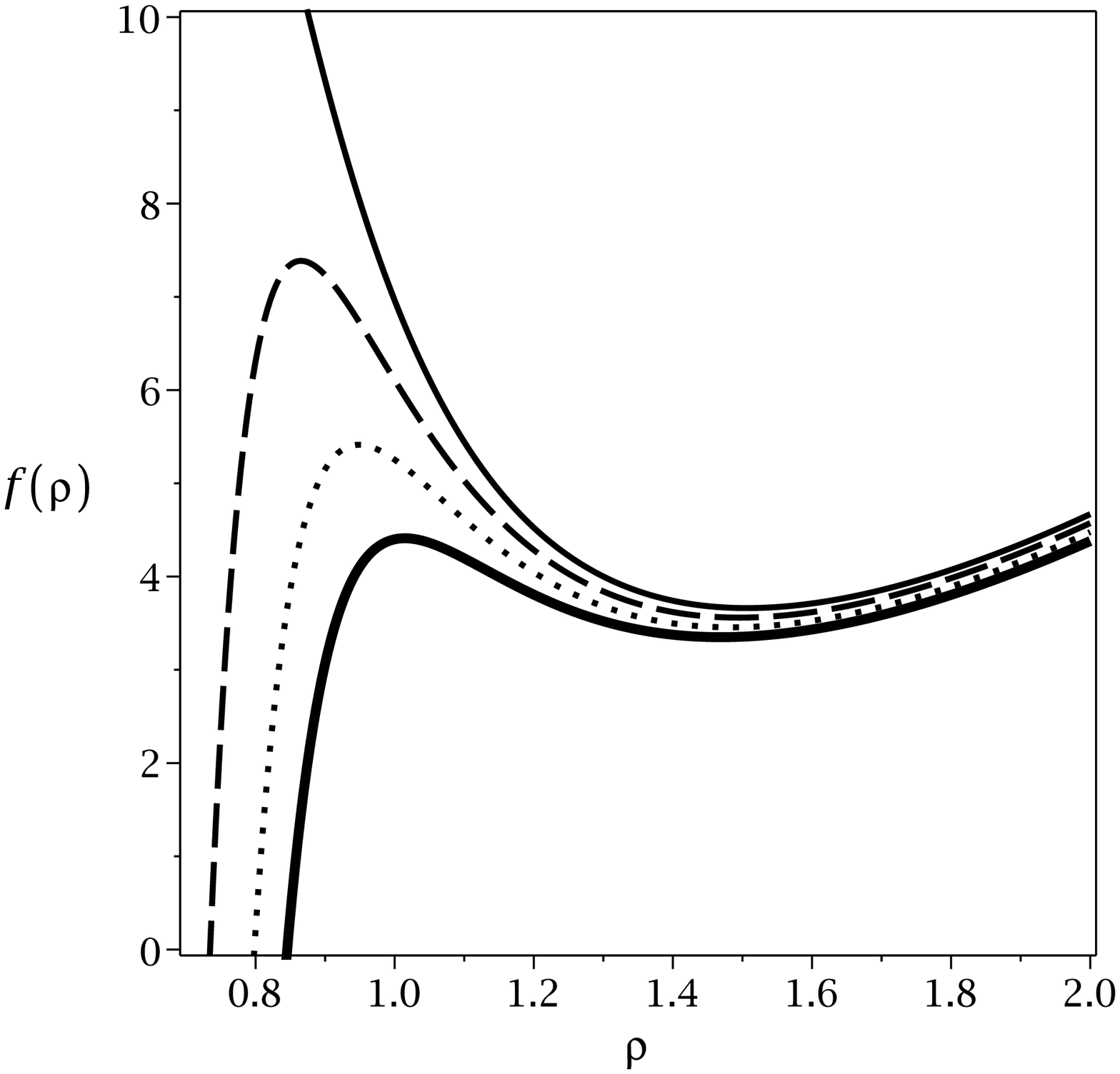} & %
\epsfxsize=5.6cm\epsffile{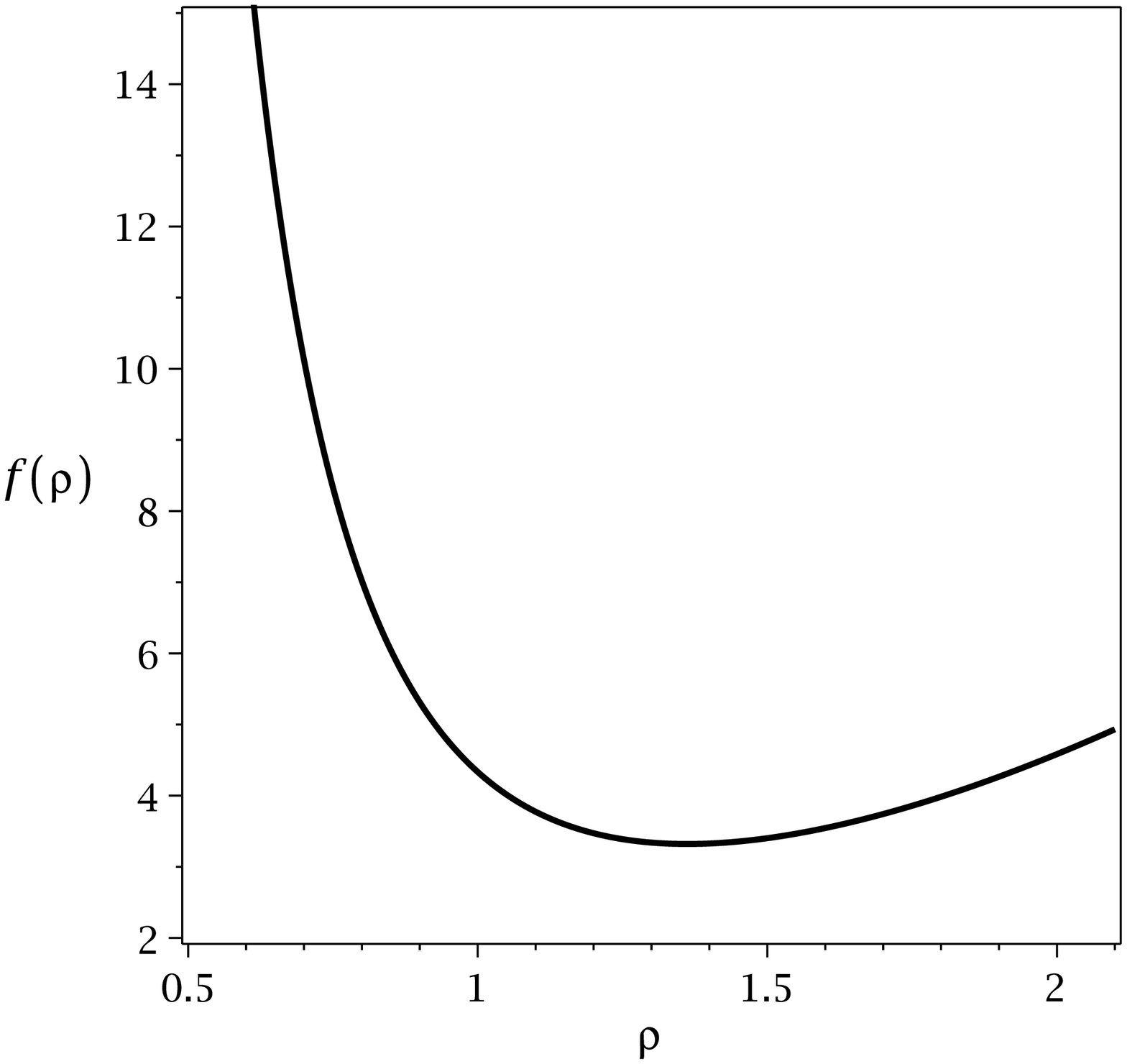}%
\end{array}
$%
\caption{f($\protect\rho$) versus $\protect\rho $ for $m=0.1$,
$q=2$, $l=1 $ and $d=5$. \newline left diagram: $\protect\alpha
=0.01$, $\protect\beta =0$ (Bold line), $\protect\beta =0.015$
(dotted line), $\protect\beta =0.025$ (dashed line) and
$\protect\beta =0.030$ (continuous line). \newline middle diagram:
$\protect\beta =0.05$, $\protect\alpha =0$ (bold line),
$\protect\alpha =0.01$ (dotted line), $\protect\alpha =0.02$
(dashed line) and $\protect\alpha =0.03$ (continuous line).
\newline right diagram: $\protect\beta=0$ and $\protect\alpha=0$.}
\label{Fig1GB}
\end{figure}

In order to obtain the metric function, $f(\rho )$, one should solve all
components of the gravitational field equation (\ref{FE2}), simultaneously.
After cumbersome calculations, we find that there are two different
differential equations with the following explicit forms
\begin{equation}
e_{\rho \rho }=\mathcal{K}_{1}+\alpha \mathcal{K}_{2}=0,  \label{er}
\end{equation}%
\begin{equation}
e_{tt}=\mathcal{K}_{11}+\alpha \mathcal{K}_{22}=0,  \label{et}
\end{equation}%
where%
\begin{eqnarray*}
\mathcal{K}_{1} &=&\Lambda +\frac{(d-2)(d-3)}{2\rho ^{2}}f+\frac{4q^{2}l^{2}%
}{\rho ^{2d-4}}+\frac{(d-2)f^{\prime }}{2\rho }-\frac{32l^{4}q^{4}\beta }{%
\rho ^{4d-8}}+O(\beta ^{2}), \\
\mathcal{K}_{2} &=&-\frac{(d-2)(d-3)(d-4)ff^{\prime }}{\rho ^{3}}-\frac{%
(d-2)(d-3)(d-4)(d-5)f^{2}}{2\rho ^{4}}, \\
\mathcal{K}_{11} &=&-\Lambda -\frac{(d-3)(d-4)f}{2\rho ^{2}}-\frac{%
(d-3)f^{\prime }}{\rho }-\frac{f^{\prime \prime }}{2}+\frac{4l^{2}q^{2}}{%
\rho ^{2d-8}}-\frac{96l^{4}q^{4}\beta }{2\rho ^{4d-12}}+O(\beta ^{2}), \\
\mathcal{K}_{22} &=&\frac{(d-3)(d-4)(d-5)}{2\rho ^{4}}\left(
(d-6)f^{2}+4\rho ff^{\prime \prime }+\frac{2\rho ^{2}}{d-5}\left( f^{\prime
}{}^{2}+ff^{\prime \prime }\right) \right) ,
\end{eqnarray*}%
Eqs. (\ref{er}) and (\ref{et}) are corresponding to $\rho \rho $ and $tt$
components of the gravitational field equation (\ref{FE2}). It is easy to
show that $\phi \phi $ and $x_{i}x_{i}$ components of Eq. (\ref{FE2}) are,
respectively, similar to $e_{\rho \rho }$ and $e_{tt}$, and therefore, it is
sufficient to solve Eqs. (\ref{er}) and (\ref{et}), simultaneously. Since we
desire to obtain higher dimensional magnetic brane solutions with GB as a
correction for EN gravity and quadratic Maxwell invariant as a correction to
the Maxwell theory, we ignored $\alpha \beta $, $\alpha ^{2}$ and $\beta
^{2} $\ terms and higher orders. Interestingly, the results for
consideration of these two corrections will be as follow
\begin{equation}
f(\rho )=f_{EN}-\frac{64q^{4}l^{4}}{\left( d-2\right) \left( 3d-7\right)
\rho ^{4d-10}}\beta +\frac{\left( d-3\right) \left( d-4\right) f_{EN}^{2}}{%
\rho ^{2}}\alpha +O\left( \alpha \beta ,\alpha ^{2},\beta ^{2}\right) ,
\label{fGB}
\end{equation}%
with%
\begin{equation}
f_{EN}=\frac{2ml^{3}}{\rho ^{d-3}}-\frac{2\Lambda }{\left( d-1\right) \left(
d-2\right) }\rho ^{2}+\frac{8q^{2}l^{2}}{\left( d-2\right) \left( d-3\right)
\rho ^{2d-6}},  \label{fEN}
\end{equation}%
where $m$ is an integration constant related to the mass. As one
can see for the case of $\alpha =\beta =0$, the effects of
corrections are cancelled and obtained results will be magnetic
solutions of EN gravity. In order to study the effects of these
two corrections on the obtained metric function, we plot some
graphs in the presence (absence) of these two corrections (see
Fig. \ref{Fig1GB}).

As one can see, in the absence of Maxwell and GB corrections, the
plotted graph for metric function versus $\rho $ is quite
different comparing to when one of the considered corrections (GB
or Maxwell) is present. Considering at least one of these
corrections will modify the behavior of metric function. This
modification is more evident and stronger for small values of
$\rho $. The metric function will have root(s) for specific value
of $\rho $ (namely $\rho _{0}$). The function $f\left( \rho
\right) $ has two extrema at $\rho _{ext_{1}}$ and $\rho
_{ext_{2}}$\ ($\rho _{ext_{1}}<\rho _{ext_{2}}$) in which for
$\rho _{0}\leq \rho \leq \rho _{ext_{1}}$, metric function is an
increasing function of $\rho $. For $\rho _{ext_{1}}\leq \rho \leq
\rho _{ext_{2}}$, $f\left( \rho \right) $ is a decreasing function
of $\rho $\ and finally, in case of $\rho \geq \rho
_{ext_{2}}$, it is an increasing function of radial coordinate (Fig. \ref%
{Fig1GB}; left and middle). The root of $f(\rho )$ is an
increasing (or decreasing) function of $\beta$ (or $\alpha$) parameter (Fig. \ref%
{Fig1GB}; left and middle). $\rho _{ext_{1}}$ is decreasing
functions of GB parameter and $f(\rho _{ext_{1}})$ is increasing
function of functions of GB parameter. As for nonlinearity
parameter, $\rho _{ext_{1}}$ is an increasing function of $\beta $
but interestingly $f(\rho _{ext_{1}})$ is a decreasing function of
it. It is worthwhile to mention that variation of $\alpha$ and
$\beta$ do not have reasonable effect on $\rho _{ext_{2}}$ and
$f(\rho _{ext_{2}})$. It is notable that for small values of the
nonlinearity parameter the metric function has no root. Contrary
to this effect, for large values of GB parameter, metric function
will be without any root. It simply shows the fact that GB and
nonlinearity parameters have opposite effects on the behavior of
metric function.

Next, we are going to discuss the geometric properties of the solutions. To
do this, we look for possible black hole solutions with obtaining the
curvature singularities and their horizons. We usually calculate the
Kretschmann scalar, $R_{\alpha \beta \gamma \delta }R^{\alpha \beta \gamma
\delta }$, to achieve essential singularity. Considering the mentioned
spacetime, it is easy to show that
\begin{equation}
R_{\alpha \beta \gamma \delta }R^{\alpha \beta \gamma \delta }=f^{\prime
\prime 2}+2\left( d-2\right) \left( \frac{f^{\prime }}{\rho }\right)
^{2}+2\left( d-2\right) \left( d-3\right) \left( \frac{f}{\rho ^{2}}\right)
^{2}.  \label{RR}
\end{equation}

Inserting the metric function, $f(\rho )$, in Eq. (\ref{RR}) and using
numerical analysis, one finds that the Kretschmann scalar diverges at $\rho
=0$ and it is finite for $\rho >0$ and naturally one may think that there is
a curvature singularity located at $\rho =0$. In what follows, we state an
important point, in which confirms that the spacetime never reaches $\rho =0$%
. As one can confirm, easily, the metric function has positive value for $%
\rho >r_{+}$. So two cases may occurr. For the first case, $f(\rho )$ is a
positive definite function with no root and therefore the singularity is
called a naked singularity which we are not interested in. We consider the
second case, in which the metric function has one or more real positive
root(s). We denote $r_{+}$ as the largest real positive root of $f(\rho )$.
The metric function is negative for $\rho <r_{+}$ and positive for $\rho
>r_{+}$ and hence the metric signature may change from ($-+++++...+$) to
($---+++...+$) in the range $\rho <r_{+}$.

We should note that this situation is different from that of black
hole solutions. Considering a typical $d$-dimensional black hole
metric $ds^{2}=-f(\rho)dt^{2}+\frac{d\rho^{2}}{f(\rho)}+\rho
^{2}d\Omega^{2}$ with $(-+++++...+)$ signature. Denoting $r_{+}$
as largest real positive root of $f(\rho)$, we know that for
$\rho>r_{+}$ the metric function is positive definite and the
signature does not change. For $\rho<r_{+}$, although the
mentioned signature changes to $(+-++++...+)$, the number of
positive and negative signs remain unchange. In other words, for
entire spacetime, black hole metric has one negative (temporal
coordinates) sign and $(d-1)$ positive (spatial coordinates)
signs. In this case, the change in sign merely signifies that this
is not the right coordinate system to study the $\rho<r_{+}$
region. But for our magnetic metric, Eq. (\ref{Metric1}), the
number of positive and negative signs will change for
$\rho<r_{+}$. Taking into account this apparent change of
signature of the metric, we conclude that one cannot extend the
spacetime to $\rho <r_{+}$. In order to get rid of this incorrect
extension, one may use the following suitable transformation with
introducing a new radial coordinate $r$
\begin{equation}
\begin{array}{c}
r^{2}=\rho ^{2}-r_{+}^{2}, \\
\rho \geq r_{+}\Longleftrightarrow r\geq 0.%
\end{array}
\label{Transformation}
\end{equation}

Using the mentioned transformation with $d\rho =\frac{r}{\sqrt{%
r^{2}+r_{+}^{2}}}dr$, one finds that the metric (\ref{Metric1}) should
change to
\begin{equation}
ds^{2}=-\frac{r^{2}+r_{+}^{2}}{l^{2}}dt^{2}+\frac{r^{2}}{\left(
r^{2}+r_{+}^{2}\right) f(r)}dr^{2}+l^{2}f(r)d\phi ^{2}+\frac{r^{2}+r_{+}^{2}%
}{l^{2}}dX^{2}.  \label{Metric2}
\end{equation}

It is worthwhile to mention that with this new coordinate, the metric
function will be in the following form

\begin{equation}
f(r)=f_{EN}-\frac{64q^{4}l^{4}}{\left( d-2\right) \left( 3d-7\right) \left(
r^{2}+r_{+}^{2}\right) ^{2d-5}}\beta +\frac{\left( d-3\right) \left(
d-4\right) f_{EN}^{2}}{r^{2}+r_{+}^{2}}\alpha +O\left( \alpha \beta, \alpha
^{2},\beta ^{2}\right) .  \label{FGB2}
\end{equation}

where%
\begin{equation}
f_{EN}=\frac{2ml^{3}}{\left( r^{2}+r_{+}^{2}\right) ^{\frac{d-3}{2}}}-\frac{%
2\Lambda }{\left( d-1\right) \left( d-2\right) }\left(
r^{2}+r_{+}^{2}\right) +\frac{8q^{2}l^{2}}{\left( d-2\right) \left(
d-3\right) \left( r^{2}+r_{+}^{2}\right) ^{d-3}}.
\end{equation}

We should note that function $f(r)$ given in Eqs. (\ref{FGB2}) is a non
negative function in the whole spacetime. Although the Kretschmann scalar
does not diverge in the range $0\leq r<\infty $, one can show that there is
a conical singularity at $r=0$. One can investigate the conic geometry by
using the \textit{circumference/radius ratio}. Using the Taylor expansion,
in the vicinity of $r=0$, we find
\begin{equation}
f(r)=f(r)\left\vert _{r=0}\right. +\left( \frac{df}{dr}\left\vert
_{r=0}\right. \right) r+\frac{1}{2}\left( \frac{d^{2}f}{dr^{2}}\left\vert
_{r=0}\right. \right) r^{2}+O(r^{3})+...,
\end{equation}%
where
\begin{equation}
\left. f(r)\right\vert _{_{r=0}}=\left. \frac{df}{dr}\right\vert _{_{r=0}}=0,
\label{ffp}
\end{equation}%
\begin{equation}
\left. \frac{d^{2}f}{dr^{2}}\right\vert _{_{r=0}}\equiv f^{\prime \prime }=%
\frac{-2\Lambda }{d-2}-\frac{8l^{2}q^{2}}{(d-2)r_{+}^{2d-4}}+\frac{%
64l^{4}q^{4}}{(d-2)r_{+}^{4d-8}}\beta +O(\beta ^{2}),  \label{ddF}
\end{equation}%
and hence
\begin{equation}
\lim_{r\longrightarrow 0^{+}}\frac{1}{r}\sqrt{\frac{g_{\phi \phi }}{g_{rr}}}%
=\lim_{r\longrightarrow 0^{+}}\frac{\sqrt{r^{2}+r_{+}^{2}}lf(r)}{r^{2}}=%
\frac{lr_{+}}{2}f^{\prime \prime }\neq 1,  \label{lim1}
\end{equation}%
which confirms that as the radius $r$ tends to zero, the limit of the
\textit{circumference/radius ratio} is not $2\pi $ and, therefore, the
spacetime has a conical singularity at $r=0$. This conical singularity may
be removed if one identifies the coordinate $\phi $ with the period
\begin{equation}
\text{Period}_{\phi }=2\pi \left( \lim_{r\longrightarrow 0}\frac{1}{r}\sqrt{%
\frac{g_{\phi \phi }}{g_{rr}}}\right) ^{-1}=2\pi \left( 1-4\mu \right) ,
\end{equation}%
where $\mu $ is given by%
\begin{equation}
\mu =\frac{1}{4}\left( 1-\frac{2}{lr_{+}f^{\prime \prime }}\right) .
\label{mu}
\end{equation}

In other words, the near origin limit of the metric
(\ref{Metric2}) describes a locally flat spacetime which has a
conical singularity at $r=0$ with a deficit angle $\delta \phi
=8\pi \mu $. Using the Vilenkin procedure, one can interpret $\mu
$ as the mass per unit volume of the magnetic brane
\cite{Vilenkin1985}. It is obvious that the nonlinearity of
electrodynamics can change the value of deficit angle $\delta \phi
$. Taking into account Eqs. (\ref{ddF}) and (\ref{mu}), we can
write
\begin{equation}
\frac{d\left( \delta \phi \right) }{d\beta }=\frac{-4\pi }{lr_{+}}\frac{%
d\left( f^{\prime \prime -1}\right) }{d\beta }=\frac{256\pi l^{3}q^{4}}{%
(d-2)r_{+}^{4d-7}f^{\prime \prime 2}}>0.  \label{betaDep}
\end{equation}

Eq. (\ref{betaDep}) indicates that $\delta \phi $ is an increasing
function of $\beta $. In addition, considering Eqs. (\ref{ddF})
and (\ref{mu}), one finds deficit angle does not depend the GB
coefficient. In order to investigate the
effects of nonlinearity, $r_{+},$ $q$, $l$ and dimensionality, we plot $%
\delta \phi $ versus $\beta $ and $r_{+}$.

\begin{figure}[tbp]
$%
\begin{array}{ccc}
\epsfxsize=8cm \epsffile{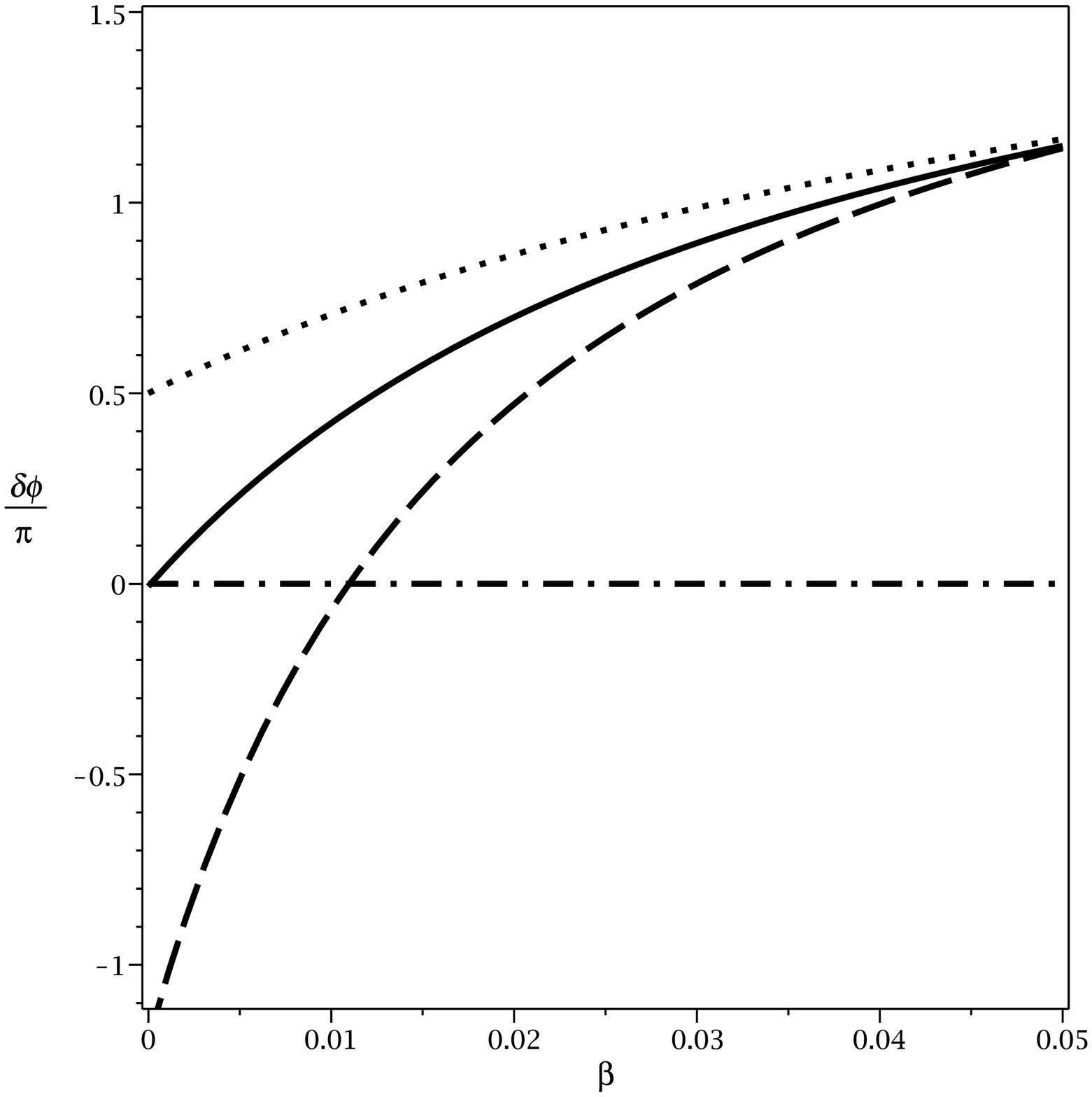} & \epsfxsize%
=8cm \epsffile{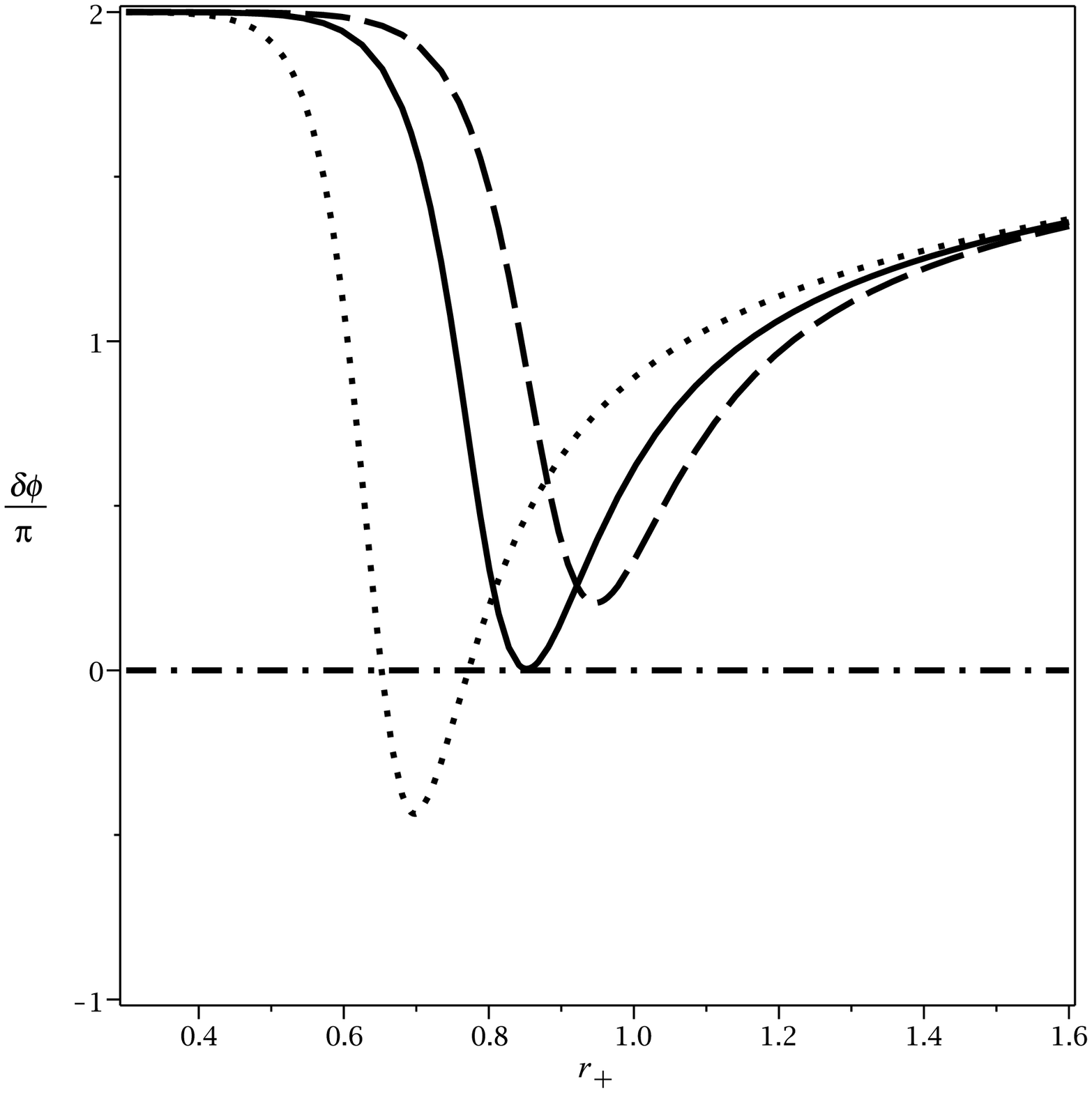} &
\end{array}
$%
\caption{$\protect\delta \protect\phi$/$\protect\pi$ versus $\protect\beta $
(left) and $\protect\delta \protect\phi$/$\protect\pi$ versus $r_{+}$
(right) for $d=5$ and $l=1$. \newline
left diagram: $r_{+}=2$, $q=8$ (dotted line), $q=8.5$ (continuous line) and $%
q=9 $ (dashed line). \newline right diagram: $\protect\beta
=0.05$, $q=0.4$ (dotted line), $q=0.73$ (continuous line) and
$q=1$ (dashed line).} \label{Fig2GB}
\end{figure}


\begin{figure}[tbp]
$%
\begin{array}{ccc}
\epsfxsize=8cm \epsffile{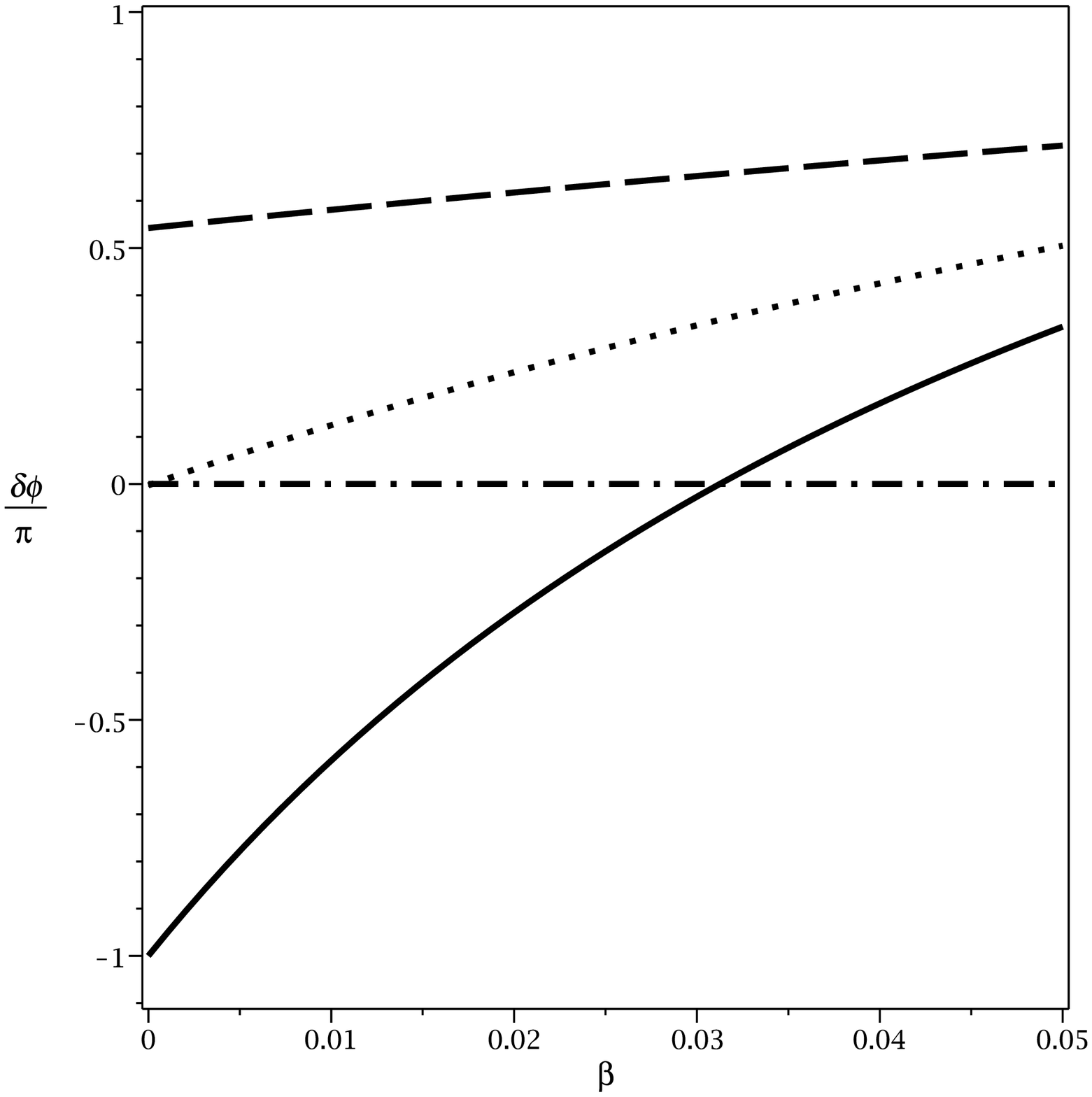} & \epsfxsize%
=8cm \epsffile{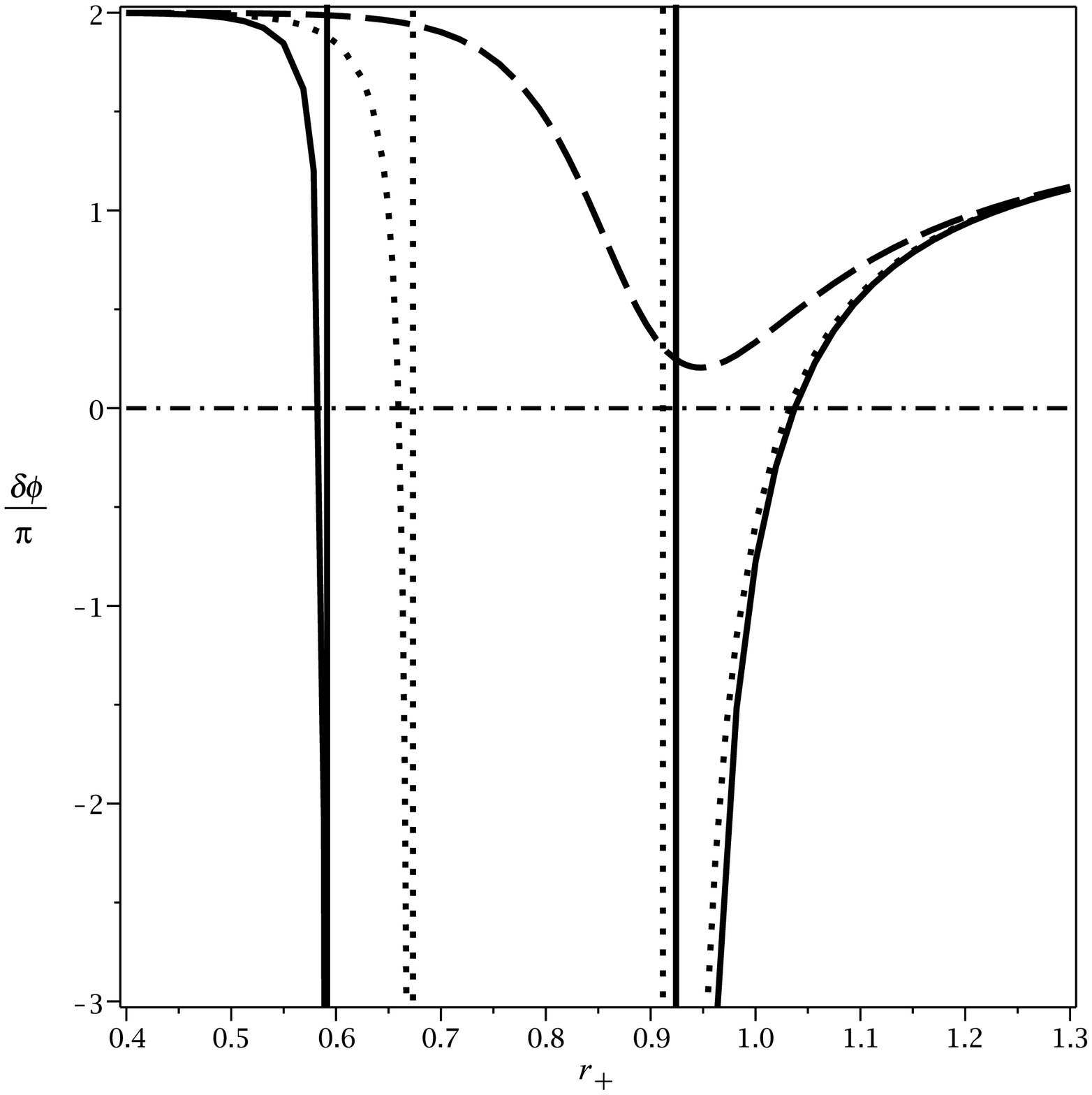} &
\end{array}
$%
\caption{$\protect\delta \protect\phi$/$\protect\pi$ versus $\protect\beta$
(left) and $\protect\delta \protect\phi$/$\protect\pi$ versus $r_{+}$
(right) for $d=5$, $l=1$ and $q=1 $. \newline
left diagram: $r_{+}=1$ (continuous line), $r_{+}=1.04$ (dotted line) and $%
r_{+}=1.1$ (dashed line), respectively. \newline right diagram:
$\protect\beta=0.005$ (continuous line), $\protect\beta=0.01$
(dotted line) and $\protect\beta=0.05$ (dashed line),
respectively.} \label{Fig3GB}
\end{figure}

\begin{figure}[tbp]
$%
\begin{array}{ccc}
\epsfxsize=8cm \epsffile{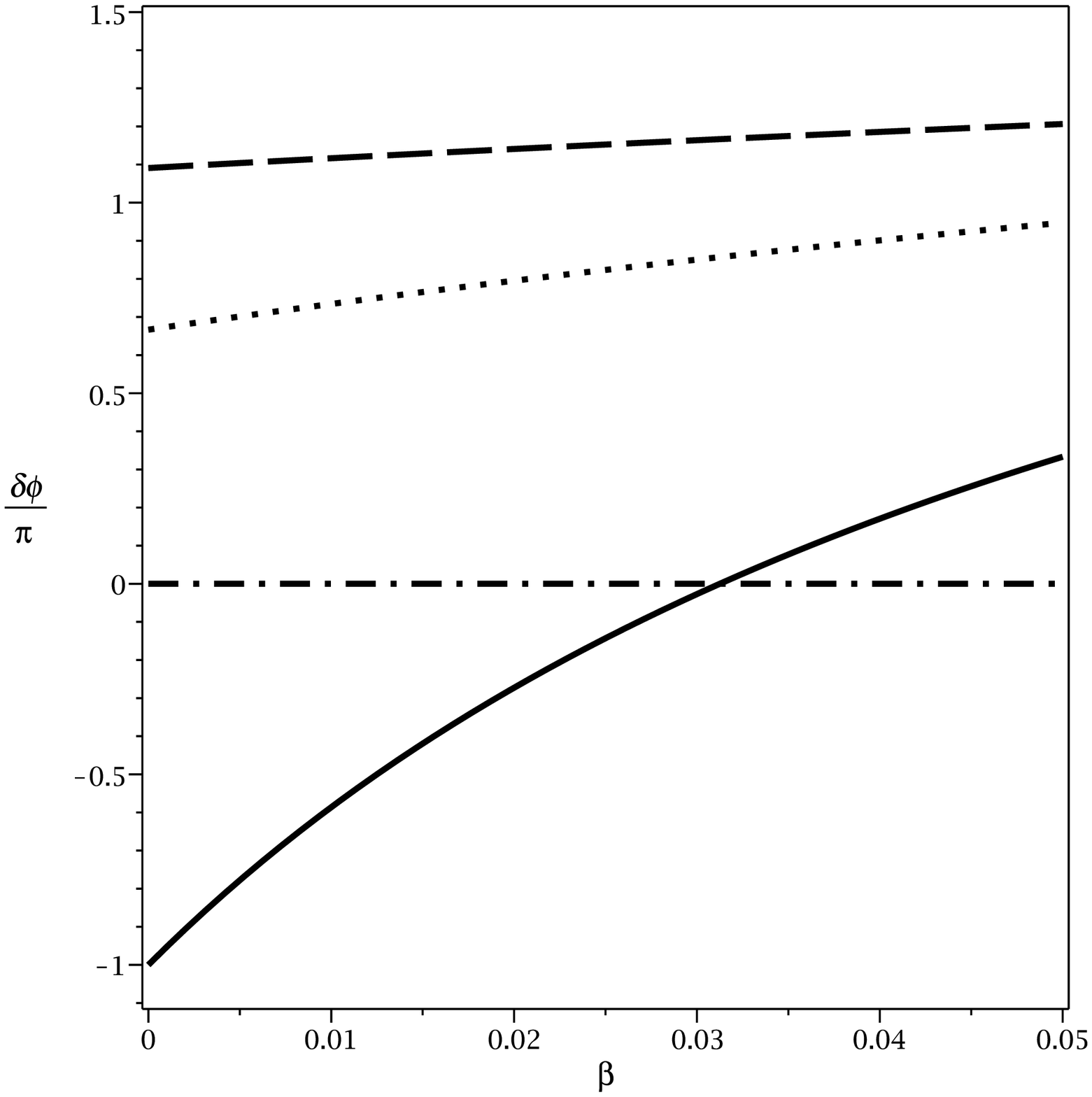} & \epsfxsize=8cm %
\epsffile{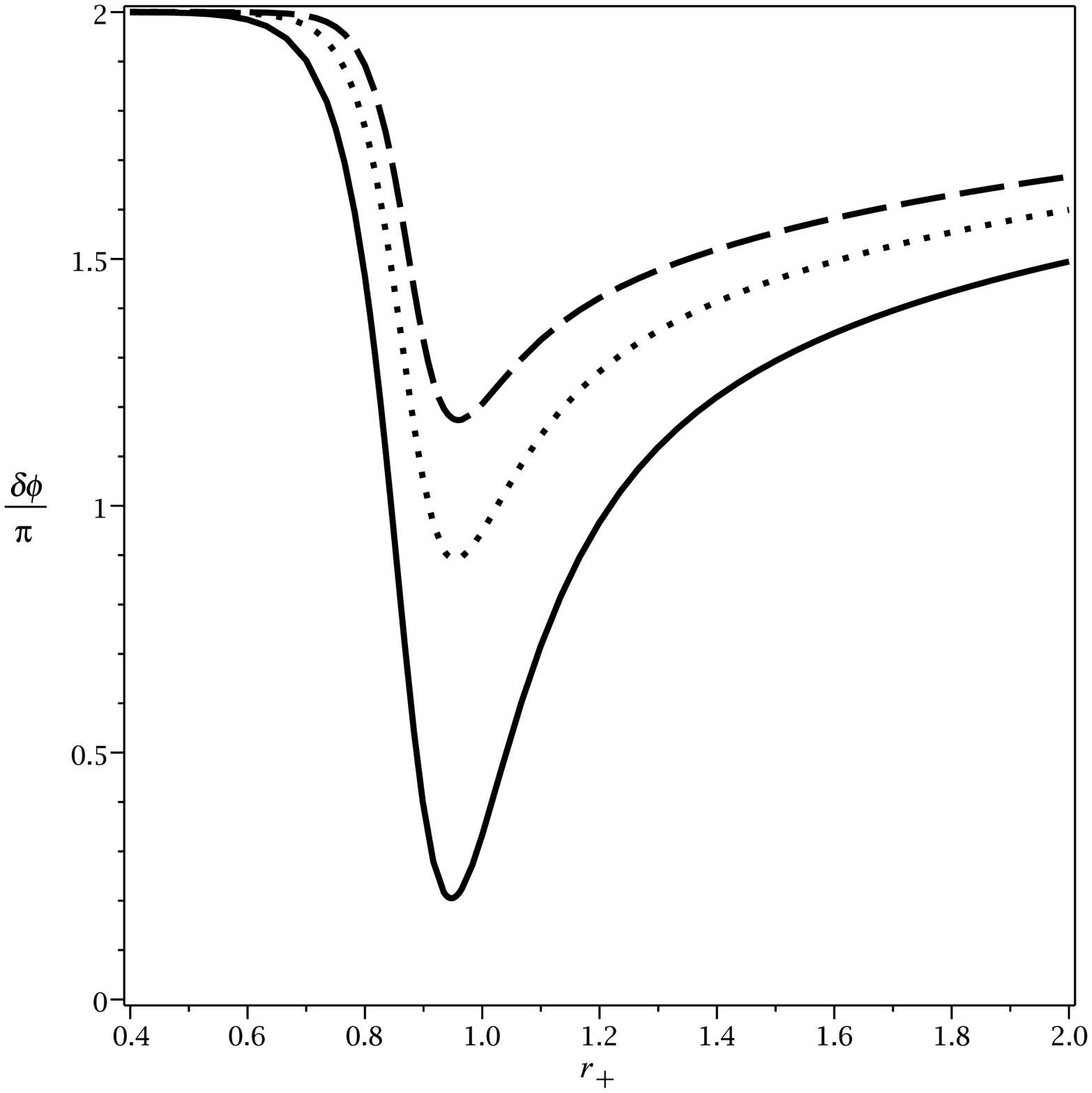} &
\end{array}
$%
\caption{$\protect\delta \protect\phi$/$\protect\pi$ versus
$\protect\beta$ (left) and $\protect\delta
\protect\phi$/$\protect\pi$ versus $r_{+}$ (right) for $q=1$ and
$l=1$. \newline left diagram: $r_{+}=1$, $d=5$ (continuous line),
$d=6$ (dotted line)and $d=7$ (dashed line). \newline right
diagram: $\protect\beta=0.05$, $d=5$ (continuous line), $d=6$
(dotted line), and $d=7$ (dashed line).} \label{Fig4GB}
\end{figure}


As one can see, deficit angle can be affected by changing the
values of $q$, $\beta $, $d$ and $r_{+}$. In order to provide
additional clarification, we plot Figs. \ref{Fig2GB}-\ref{Fig4GB}.
Considering these figures, left panels indicate the variation of
deficit angle versus $\beta $, whereas the
right panels correspond to the behavior of deficit angle with respect to $%
r_{+}$. Left panel of Figs. \ref{Fig2GB}-\ref{Fig4GB} confirm that
deficit angle is an increasing function of $\beta$. Moreover Fig.
(\ref{Fig2GB}, left) shows that for $\beta \longrightarrow 0$
(Maxwell case), deficit angle is a decreasing function of $q$. In
addition, there is a $q_{c}$, in which for $q>q_{c}$ deficit angle
is negative for $\beta =0$ (one can obtain $q_{c}$ in such a way
that $\delta \phi \left\vert _{\beta =0,q=qc}\right. =0$).

In case of deficit angle versus $r_{+}$ for different values of
charge parameter (Fig. \ref{Fig2GB}, right), the plotted graph is
divided into three distinguishable regions.  We find that
variation of $q$, $\beta$ and $q$ do not affect, significantly,
the behavior of deficit angle for sufficiently small (large)
$r_{+}$ ( see right panel of Figs. \ref{Fig2GB}-\ref{Fig4GB}). In
addition Fig. \ref{Fig2GB} (right) shows there is an extremum
point $r_{+_{ext}}$, in which for $ r_{+}\leq r_{+_{ext}}$
($r_{+_{ext}}\leq r_{+}$), deficit angle is a decreasing (an
increasing) function of $r_{+}$. The mentioned $r_{+_{ext}}$ and
its corresponding deficit angle are increasing functions of charge
parameter. In other words, for adequately small charge the minimum
value of deficit angle is negative (Fig. \ref{Fig2GB}, right). By
increasing charge the region of negativity and hence the distance
between roots of deficit angle decreases.

Now, we plot deficit angle versus $r_{+}$ for various $\beta $
(Fig. \ref{Fig3GB}, right) to show an interesting behavior. This
figure indicates two divergency for deficit angle when $\beta
<\beta _{c}$. In other words, these singularities, $r_{+_{1}}$ and
$r_{+_{2}}$ ($r_{+_{1}}<r_{+_{2}}$), can be removed by increasing
$\beta$. For $\beta >\beta _{c}$, deficit angle has a minimum and
for $\beta <\beta _{c}$, we cannot obtain physical
($\protect\delta \protect\phi \leq 2\pi$) deficit angle for
$r_{+_{1}}<r_{+}<r_{+_{2}}$. Deficit angle before first singular
point is a decreasing function of $r_{+}$ whereas, it is an
increasing function after second singular point. Variation of the
nonlinearity parameter changes the distance between these two
singular points. In other words, increasing $\beta $ leads to
decrease the mentioned interval. Fig. \ref{Fig3GB}; left),
indicates that deficit angle is an increasing function of $r_{+}$
in this parameter region.

Finally, we are considering the effects of dimensions on deficit
angle (Fig. \ref{Fig4GB}). As one can see, for certain dimensions,
there is a region for $\beta$ where, the deficit angle is negative
(Fig. \ref{Fig4GB} left) and increasing dimensions leads to
vanishing this region. Fig. \ref{Fig4GB} shows that deficit angle
is an increasing function of dimensions in this parameter region.
As for deficit angle versus $r_{+}$, (Fig. \ref{Fig4GB} right),
there is a similar behavior with (Fig. \ref{Fig2GB} right) and
deficit angle has a minimum.

In order to explain the negative deficit angle, we first describe positive
one. Cutting segment of a certain angular size of a two dimensional plain
and then sewing together the edges to obtain a conical surface. This conical
space is flat but has a singular point corresponding to the apex of the
cone. The deleted segment from the plan is known as deficit angle with
positive values. According to the previous statement, here, we imagine a new
situation when a segment is added to the new plane to obtain a flat surface
with a saddle-like cone (for more details see Fig. $2$ in Ref. \cite{LR}).
This added segment is corresponding to a negative deficit angle (or surplus
angle).

It is worthwhile to mention that although the deleted segment is bounded by
the value of $2\pi$ the added segment is unbounded. Therefore, one can
conclude that the range of deficit angles is from $-\infty$ to $2\pi$.
Positive/negative deficit angles may be related to the positive/negative
torsion of space or the attractive-type/repulsive-type of gravitational
potentials and more details of negative deficit angle with its physical
interpretations can be found in \cite{LR,NegDA}.

\section{Closing Remarks}

In this paper, we have considered two different kinds of
corrections to both matter and gravitational fields. For
gravitational aspect, we have considered GB gravity as a
correction to EN gravity whereas for electromagnetic aspect, we
have regarded a quadratic power of Maxwell invariant in addition
to Maxwell Lagrangian as a correction to electromagnetic field.
Remarkably it was seen that, in absence of these two corrections,
the behavior of the metric function is completely different
comparing to consideration of at least one of them.

Interestingly, the root(s) of metric function was also modified by
considering these corrections. The place of this root(s) was a
decreasing (an increasing) function of GB (nonlinearity)
parameter. For small values of the nonlinearity parameter and
large values of the GB parameter, the behavior of the metric
function was similar to the case of Einstein-Maxwell. In addition,
we found that the contribution of considered matter field was
opposite of the gravitational field. We found that for large
values of radial coordinate these corrections do not have
significant effect on the metric function. In other words, the
dominant region in which these two corrections modify metric
function meaningfully was for small values of $\rho$.

Next, in order to avoid change of signature, we used a suitable
radial transformation and found that there is no curvature
singularity through the whole of spacetime, but there is a conical
singularity located at the origin. We have studied the behavior of
deficit angle and the effects of different parameters on it. We
have plotted two kinds of diagrams. One is
for deficit angle versus $\beta$ and the other one is deficit angle versus $%
r_{+}$. For the case of deficit angle versus $\beta $, due to fact
that we considered nonlinearity as a correction, we only have
taken small values of it into account. In general the behavior of
the left graphs were monotonic and the calculated deficit angles
were increasing function of nonlinearity parameter. As for deficit
angle versus $r_{+}$, interestingly, the behavior was completely
different to the other case. In this case, for variation of $q$
and $d$, we found that for sufficiently small or large $r_{+}$
deficit angle is independent of the value of other parameters. In
addition, we showed that there is a minimum value for deficit
angle in a specific $r_{+}$. Moreover, we found that the only
parameter that modified the general behavior of these graphs,
significantly, were nonlinearity parameter. For small values of
$\beta$ two singular points were seen. Interestingly, as one
increases the nonlinearity parameter the distance between these
two singular points decreases. In other word, by increasing
nonlinearity parameter, a compactification happens which decreases
the region between two singular points to a level that the
mentioned singularities vanish. This behavior shows the fact that
small values of $\beta$ have stronger contribution comparing to
other parameters drastically.

The existence of root, region of negativity and divergency for
deficit angle are other important issues that must be taken into
account. In studying deficit angle, we are considering second
order derivation of metric function with respect to radial
coordinate. Considering the fact that metric function could be
interpreted as a potential (see for example chapter $9$ of Ref.
\cite{dInverno}), it is arguable to state that the singular point
may be indicated as a phase transition. On the other hand,
geometrical structure of the solutions in case of positive and
negative values of deficit angle is different. In case of positive
deficit angle, the geometrical structure of the object is cone
like with a deficit angle whereas in case of the negative angle
the structure will be saddle-like with surplus angle. We found
that for positive deficit angle, there is an upper limit whereas
for negative deficit angle there is no limit. Therefore, one may
state that due to these differences in the structure of the
solutions, the root of deficit angle may represent a phase
transition. These two arguments could be discussed in more details
if the physical concept of negative deficit angle have become more
clear. Moreover, we should note that, spacetime has no deficit
angle for vanishing $\delta \phi $. In other words, the
geometrical structure of the solutions in this case represents no
defect. Therefore, one may argue that these cases are representing
the magnetic brane without conical structure.

Finally, we will be interested in analyzing the theory of gravity
that was proposed in this paper in more details and calculating
conserved quantities of this theory. Also one can consider higher
orders of the Lovelock gravity as corrections to the EN gravity
and study their effects on magnetic branes and their deficit
angle. Phase transition, structural properties and physical
behavior of different objects have been studied through these
defects. Since obtained solutions are asymptotically AdS, it is
worthwhile to consider these solutions in context of AdS/CFT
correspondence and study different phenomena through these
solutions. In addition, one can make some modifications regarding
the geometry of the solutions and remove the mentioned conical
singularity, and then, use the copy-and-paste method to obtain
geodesically complete spacetime with a minimum value for
$\rho_{min}=r_{+}$ as a throat \cite{worm}. We left these issues
for the future works.

\begin{acknowledgements}
We thank an unknown reviewer for helpful advices. We also thank
the Shiraz University Research Council. This work has been
supported financially by the Research Institute for Astronomy and
Astrophysics of Maragha, Iran.
\end{acknowledgements}

\end{document}